\documentclass[]{elsarticle}
\usepackage{lineno,hyperref}
\usepackage{epstopdf} 
\usepackage{color}
\usepackage{lscape}
\modulolinenumbers[5]
\biboptions{sort&compress}
\usepackage{graphicx}
\usepackage{amssymb}
\begin{document}
\begin{frontmatter}
\title{Defect-induced ordering and disordering in metallic glasses}

\author[VSPU]{A.S. Makarov}
\author[VSPU]{G.V. Afonin}
\author[VSPU]{R.A. Konchakov}
\author[NWPU]{J.C. Qiao}
\author[MISiS]{A.N. Vasiliev}
\author[IFTT]{N.P. Kobelev}
\author[VSPU]{V.A. Khonik}
\corref{cor}
\cortext[cor]{Corresponding author} 
\ead{v.a.khonik@yandex.ru} 
\address[VSPU] {Department of General Physics, Voronezh State Pedagogical
University,  Lenin St. 86, Voronezh 394043, Russia}
\address[NWPU] {School of Mechanics, Civil Engineering and Architecture, Northwestern Polytechnical University, Xi’an 710072, China}
\address[MISiS] {National University of Science and Technology MISiS, Moscow 119049, Russia}
\address[IFTT] {Institute of Solid State Physics RAS, Chernogolovka, 142432, Russia}

\begin{abstract}

On the basis of shear modulus measurements on a Pt-based glass, we calculated temperature dependence of the defect concentration $c$ using the Interstitialcy theory. This temperature dependence is compared with temperature dependence of the normalized full width at half maximum (FWHM) $\gamma$ of the first peak of the structure factor $S(q)$ for the same glass available in the literature.  It is found that $\gamma$ above the glass transition temperature $T_g$ linearly increases with $c$ in the same way for both initial and relaxed (preannealed) samples providing the evidence of defect-induced disordering in the supercooled liquid region independent of glass thermal prehistory. For both states of the samples, the derivative $d\gamma/dc$ is  close to unity. Below $T_g$, the interrelation between $\gamma$ and $c$ is entirely different for initial and relaxed samples. In the former case, strong defect-induced ordering upon approaching $T_g$ is observed while relaxed samples do not reveal any clear ordering/disordering. Possible reasons for these observations are discussed.

To further investigate the  relationship between the normalized FWHM and defect concentration, we  performed molecular dynamic simulation of $\gamma (c)$-dependence in a high-entropy FeNiCrCoCu  model glass. It is found that $\gamma$ also linearly increases with $c$ while the derivative $d\gamma /dc$ is  again close to unity just as in the case of Pt-based glass. 

\end{abstract}


\end{frontmatter}

\section{Introduction}

Major information on the structure of metallic glasses  (MGs) is derived most often from the structure factor $S(q)$, which is calculated from primary diffraction data  \cite{ChengProgMatSci2011}. In particular, the non-crystalline structure of MGs can be characterized by the full width at half maximum (FWHM) of the first $S(q)$-peak, which constitutes an integral measure of structural disordering. It has been found that the FWHM is quite sensitive to changes of different experimental conditions. Specifically, while the  position of the first $S(q)$-peak significantly varies with temperature and can be used for  the estimates of  the volume thermal expansion and density changes upon structural relaxation, the FWHM rapidly increases with temperature above the glass transition temperature $T_g$ (i.e. in the supercooled liquid state)  \cite{NeuberActaMater2021,YavariActaMater2005,GeorgarakisActaMater2015,SinghJALCOM2015,DuMaterToday2020,LuanNatCommun2022}. The FWHM also rises with the melt quenching rate and shows that melt-spun ribbon MGs are more disordered as compared with bulk MGs produced by melt suction \cite{SinghJALCOM2015}. On the other hand, an increase of the melt quenching rate leads to a decrease of the hardness and changes the wear performance \cite{ChenMaterSciEng2008,HuangMaterDes2014}. Plastic deformation by cold rolling or  high-pressure torsion significantly increase the FWHM indicating intense structural disordering \cite{XuMaterSciEng2015,AdachiMaterrSciEng2015,CaoActaMater2006,EbnerActaMater2018,GunderovMaterLett2020}.
An important information was derived by the authors \cite{LiNatureMater2022} who fabricated over five thousand alloys and found a strong correlation between the high glass forming ability and  FWHM showing  that a large dispersion of structural units comprising the amorphous structure is a universal indicator for the high metallic glass-formating ability \cite{LiNatureMater2022}. Thus, the FWHM constitutes an important integral   parameter showing the degree of MGs' structural disordering and sensitive to different treatments.

In this work, we are interested in temperature impact on the FWHM at temperatures  both below and above $T_g$. In the latter case, as mentioned above, the FWHM rapidly increases with temperature and it is suggested that this effect is due to increasing atomic vibrations \cite{NeuberActaMater2021}. However, a simple increase of atomic vibrations should provide a monotonous FWHM rise independent of any specific temperature range under consideration, which is clearly not the case. Therefore, the nature of rapid broadening of the structural factor in the supercooled liquid state above $T_g$ deserves closer attention.  

Currently, a widely accepted idea is that the structure of MGs can be represented as  a non-crystalline matrix with  dominant (usually icosahedral) packing and regions with low point symmetry commonly called defects (e.g. Refs \cite{ChengProgMatSci2011,HirataScience2013}). In this case, an increase of the defect concentration should increase the FWHM due to the rise of the amount of regions with damaged dominant short-range order. To further utilize this idea, one has to accept a specific theory (model) of defects in MGs. In our opinion, the Interstitialcy theory  (IT) is suitable for this purpose. By using the IT, we calculated the defect concentration below and above $T_g$ and showed that in the latter case this concentration is directly proportional to the experimental FWHM for a particular Pt-based metallic glass. This fact indicates that it is the defect concentration, which determines a rapid rise of the FWHM at  $T>T_g$. Below $T_g$, the situation is more complicated and the FWHM depends on the state (initial/relaxed) of glass.

To further study the relationship between the FWHM and defects, we performed molecular dynamic simulations of a model metallic glass produced by quenching the melt at various rates and containing, therefore, different amount of defects in the solid glassy state. We calculated the model structure factor and found that the derivative of the FWHM over the defect concentration is quite close to the one determined from the analysis of the experimental data on Pt-based glass. This provides further argument that the broadening of the structural factor of MGs is directly related to the defect subsystem.

\section{Experimental and modeling procedures }           

\subsection{Experimental} The experimental part of the present investigation was performed on bulk glassy Pt$_{42.5}$Cu$_{27}$Ni$_{9.5}$P$_{21}$ (at.\%). The reason for this choice is determined by detailed \textit{in situ} measurements of the FWHM for this glass in the course of heating available in the literature \cite{NeuberActaMater2021}. The master alloy of the above composition was produced by direct fusing of the constituent elements (at least 99.9\% pure) in evacuated quartz vial using a two-temperature method. The glass was next produced by a melt jet quenching method. The castings were verified to be X-ray amorphous using a Siemens D-500 diffractometer operating in Co$_{K_{\alpha}}$ radiation.  The density of the glass was found to be 13.3$\pm{0.1}$ g/cm$^3$. Room-temperature elastic constants, shear modulus $G_{rt}=32.7$ GPa, Young's modulus $E_{rt}=92.9$ GPa, bulk modulus $B_{rt}=198.7$ GPa and Poisson ratio $\nu=0.42$,  were determined by the resonant ultrasound spectroscopy using the setup similar to that described in Ref.\cite{BalakirevRevSciInstrum2019}. 

The defect concentration was determined on the basis of high-frequency shear modulus measurements. For this, the electromagnetic acoustic transformation (EMAT) method was used \cite{VasilievUFN1983}. In this method, the Lorentz interaction between the current induced in sample's surface layer by an exciting coil and external magnetic field is used to produce resonant vibrations. For this purpose, frequency scanning was automatically performed every 10-15 s upon heating (3 K/min) and the transverse resonant frequency $f$ of a 5$\times$5$\times$2 mm$^3$ specimen was determined as a maximal signal response received by a pick-up coil upon scanning in a vacuum $\approx{0.01}$ Pa. The shear modulus was then calculated as  $G(T)=G_{rt}f^2(T)/f^2_{rt}$, where $f_{rt}$ is the resonant frequency (450-550 kHz) at room temperature. The error in the measurements of $G(T)$-changes was estimated to be $\approx 5$ ppm near room temperature and about 100 ppm near the glass transition temperature $T_g$. 

Differential scanning calorimetry (DSC) was carried out by a Hitachi DSC 7020 instrument operating in high purity (99.999 \%) nitrogen atmosphere at a rate of 3 K/min using 120-130 mg samples. Each DSC run was carried out with a fully crystallized sample of the same composition and about the same mass in the reference cell.

The defect concentration in Pt$_{42.5}$Cu$_{27}$Ni$_{9.5}$P$_{21}$ glass was calculated within the framework of the Interstitialcy theory (see below). For this, one needs to know the shear modulus $\mu$ of the maternal crystalline state. This does not pose a problem for an analysis of experimental data but represents the difficulty in computer simulation. Molecular dynamic simulation of the above glass is not feasible due to the absence of the information on the interatomic interaction in this system. Thus, the selection of a model system constitutes a problem. In particular, the choice of a very popular Zr-Cu system for modeling  is not acceptable because it has a number of different crystalline phases and it is unclear which of them represents the maternal crystalline state. In the view of the above, we performed computer modeling  on high-entropy Fe$_{20}$Ni$_{20}$Cr$_{20}$Co$_{20}$Cu$_{20}$ system (at.\%) because it crystallizes into a single FCC phase with easily determinable shear modulus. It is this phase, which was accepted as the maternal crystalline state.    

\subsection{Modeling details}
Molecular dynamic simulation was performed using the LAMMPS package \cite{PlimptonJCompPys1995}. The model size was accepted to be 32 000 atoms (20$\times$20$\times$20 translations of the FCC lattice in the crystalline state). Many-body embedded atom  potential was taken from the work \cite{FarkasJMaterRes2018}. This potential was earlier used for an analysis of points defects in the single crystalline state  and defect identification in the glassy state of this alloy \cite{KretovaJETPLett2020,KonchakovJETPLett2022}. The initial single-crystalline system was obtained by random distribution of atoms in the FCC lattice nodes with the preservation of the nominal chemical composition. The system was next melted by heating up to 3000 K and quenched to zero temperature at different rates ranging from $6\times 10^{12}$ K/s  to $1\times 10^{14}$ K/s as indicated below. This allowed to obtain glassy states with different degree of relaxation, which can be rationalized in terms of different concentration of frozen-in defects. The structure factor was calculated using the OVITO software \cite{StukowskiModelSimulMaterSciEng2010}.  

\section{Results and discussion}
\subsection{Experimental data and their analysis}

\begin{figure}[t]
\begin{center}
\includegraphics[scale=0.30]{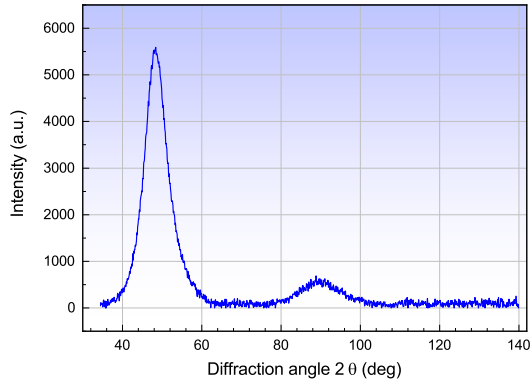}
\caption[*]{\label{Fig1.eps} X-ray diffraction of bulk glassy Pt$_{42.5}$Cu$_{27}$Ni$_{9.5}$P$_{21}$. }
\end{center}
\end{figure} 

\begin{figure}[t]
\begin{center}
\includegraphics[scale=0.6]{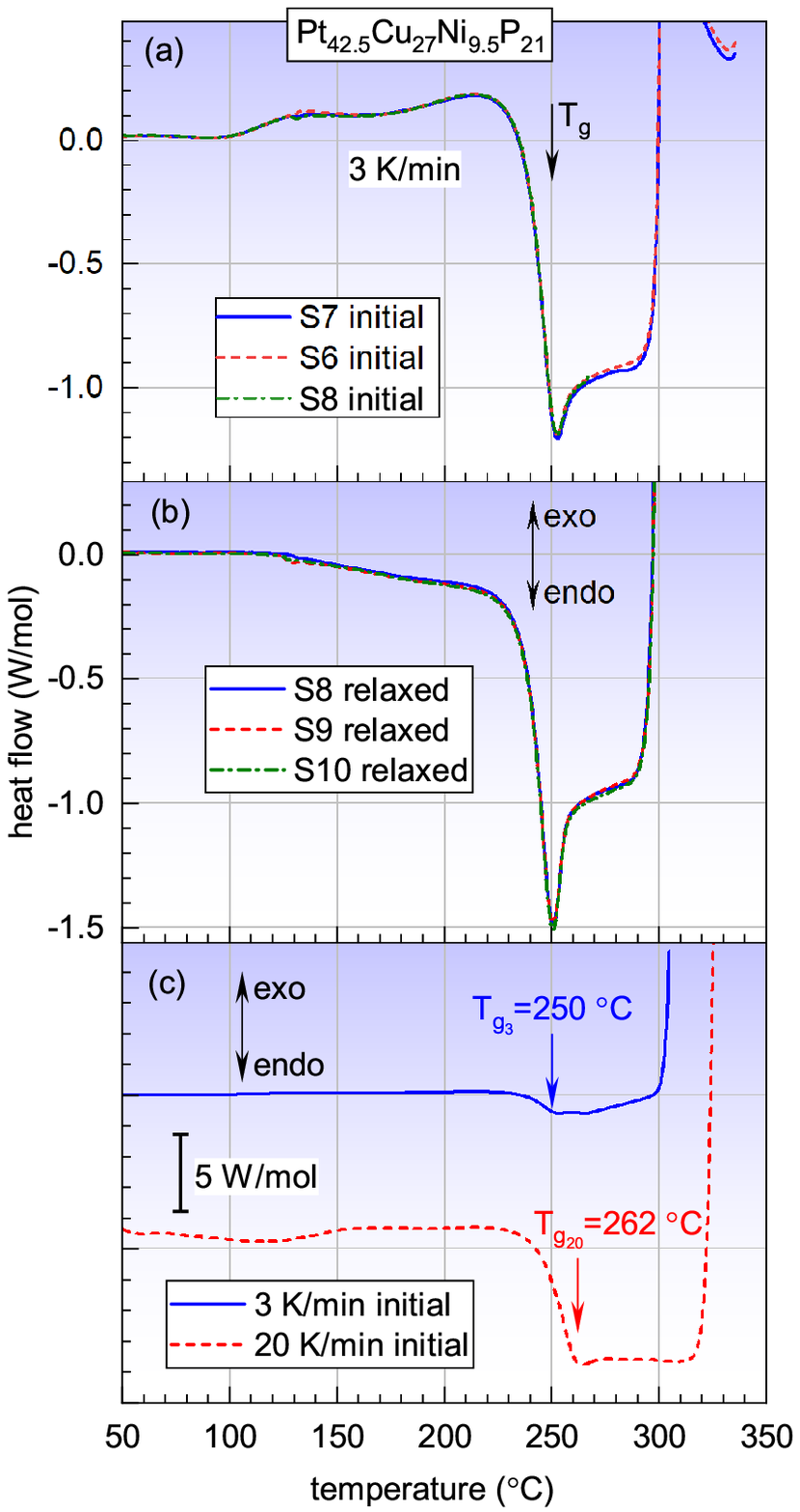}
\caption[*]{\label{Fig2.eps} DSC traces of the samples in the initial (a) and relaxed (b) states at a rate of 3 K/min. The data for three initial and three relaxed samples are shown in order to demonstrate the excellent reproducibility of the measurements. It is seen that initial samples display heat release below the glass transition temperature $T_g$ while relaxed samples exhibit only heat absorption at $T<T_g$. Panel (c) gives DSC runs for the initial samples at the heating rates of 3 K/min and 20 K/min. The corresponding $T_g$s are shown by the arrows. It is seen that the increase of the heating rate rises $T_g$ by  $\approx 12^{\circ}$C. }
\end{center}
\end{figure} 

Figure \ref{Fig1.eps} presents X-ray diffraction of the glass under investigation, which is a typical non-crystalline pattern without any signs of crystallinity. Figure \ref{Fig2.eps} gives DSC traces of samples in the initial state (a) and after relaxation (b) performed by heating up to a temperature $T=286$ $^{\circ}$C, which is deep in the supercooled liquid state but below the crystallization onset temperature. It is seen that initial samples display exothermal reaction below $T_g$ (shown by the arrows) while relaxed specimens exhibit endothermal reaction at $T<T_g$. Thermal behavior of initial and relaxed samples above $T_g$ is generally similar. Panel (c) in Fig.\ref{Fig2.eps} gives DSC traces of initial samples, which for the purpose of the present investigation were taken at two different heating rates, 3 K/min and 20 K/min. It is seen that the above increase of the heating rate increases $T_g$ by about 12$^{\circ}$C.  

\begin{figure}[h]
\begin{center}
\includegraphics[scale=0.35]{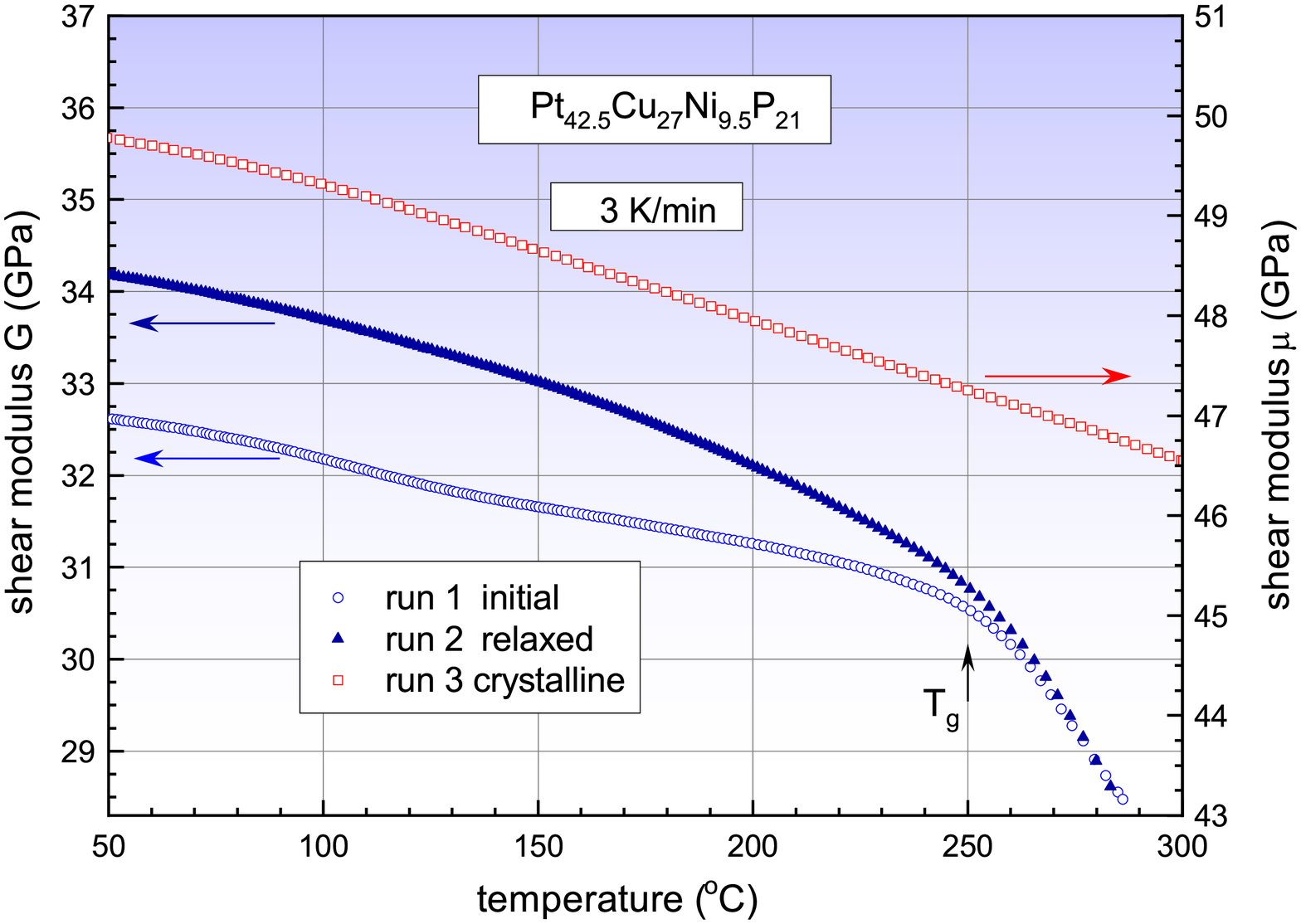}
\caption[*]{\label{Fig3.eps} Temperature dependences of the shear modulus in the initial, relaxed and crystalline states. Calorimetric glass transition temperature $T_g$ is indicated.}
\end{center}
\end{figure} 

The goal of the present paper is to relate the FWHM with the defect concentration as a function of temperature. This can be done using the aforementioned Interstitialcy theory (IT), which provides comprehensive understanding of different relaxation phenomena in MGs (see e.g. Refs \cite{KhonikMetals2019,MakarovJPCM2021,MakarovJETPLett2022} and papers cited therein).  
In particular, the IT provides almost exact description of the kinetics of heat effects on the basis of shear modulus relaxation data and gives adequate description of all thermodynamic excess potentials of MGs \cite{KhonikMetals2019,MakarovJETPLett2022}. 
Within the IT framework, temperature dependence of the concentration $c$ of interstitial-type defects is related to temperature dependences of the unrelaxed (high-frequency) shear moduli of glass and maternal crystal, $G$ and $\mu$, respectively, as 

\begin{equation}
c(T)=\frac{1}{\alpha \beta}ln\left[\frac{\mu(T)}{G(T)}\right],\label{c}
\end{equation}
where $\beta$ is a dimensionless shear susceptibility (of about 20 for different MGs), which is linked to defect-induced shear softening and anharmonicity of the interatomic potential \cite{MakarovIntermetallics2017}, while dimensionless $\alpha\approx 1$ is related to the defect strain field \cite{KhonikMetals2019}. 

Thus, the defect concentration can be calculated using shear modulus measurements on glassy and crystalline samples. The results of such investigation are presented in Fig.\ref{Fig3.eps}, which gives three shear modulus measurement runs taken subsequently on the same sample. Run 1 shows a monotonous decrease of $G$ up to the calorimetric glass transition temperature $T_g$ (shown by the arrow). At $T>T_g$, a pronounced shear softening (decrease of $G$) is observed. The heating was stopped at $T=286$ $^{\circ}$C (deep in the supercooled liquid state, see Fig.\ref{Fig2.eps}) and the sample was cooled back to room temperature at the same rate that results in the increase of the shear modulus by $\approx 5.2\%$. Run 2 gives the shear modulus in the relaxed state produced by the previous run. The sample was heated up to a temperature $T=333^{\circ}$C, which produces the complete crystallization and the shear modulus at room temperature  becomes by $\approx 67\%$ larger than that in the initial state. Finally, run 3 gives temperature dependence of the shear modulus $\mu$ in the crystalline state.    

To calculate the defect concentration with Eq.(\ref{c}), one needs to know the shear susceptibility  $\beta$. For this, we used the method presented in Ref.\cite{MitrofanovJALCOM2016} according to which this parameter can be calculated as

\begin{equation}
\beta=\frac{\Delta G_{rel}}{\rho Q_{rel}}, \label{beta}
\end{equation}
where $\Delta G_{rel}=G_{rel}^{rt}-G_{ini}^{rt}$ is the change of the shear modulus at room temperature due structural relaxation with $G_{rel}^{rt}$ and $G_{ini}^{rt}$ being  the shear moduli at room temperature before and after structural relaxation, respectively, $Q_{rel}$ is the corresponding heat release and $\rho$ is the density. With $G_{ini}^{rt}= 32.7$ GPa and $G_{rel}^{rt}=34.3$ GPa (see Fig.\ref{Fig3.eps}) and measured $Q_{rel}=6670$ J/kg, with Eq.(\ref{beta}) one arrives at $\beta=18$. 

\begin{figure}[t]
\begin{center}
\includegraphics[scale=0.35]{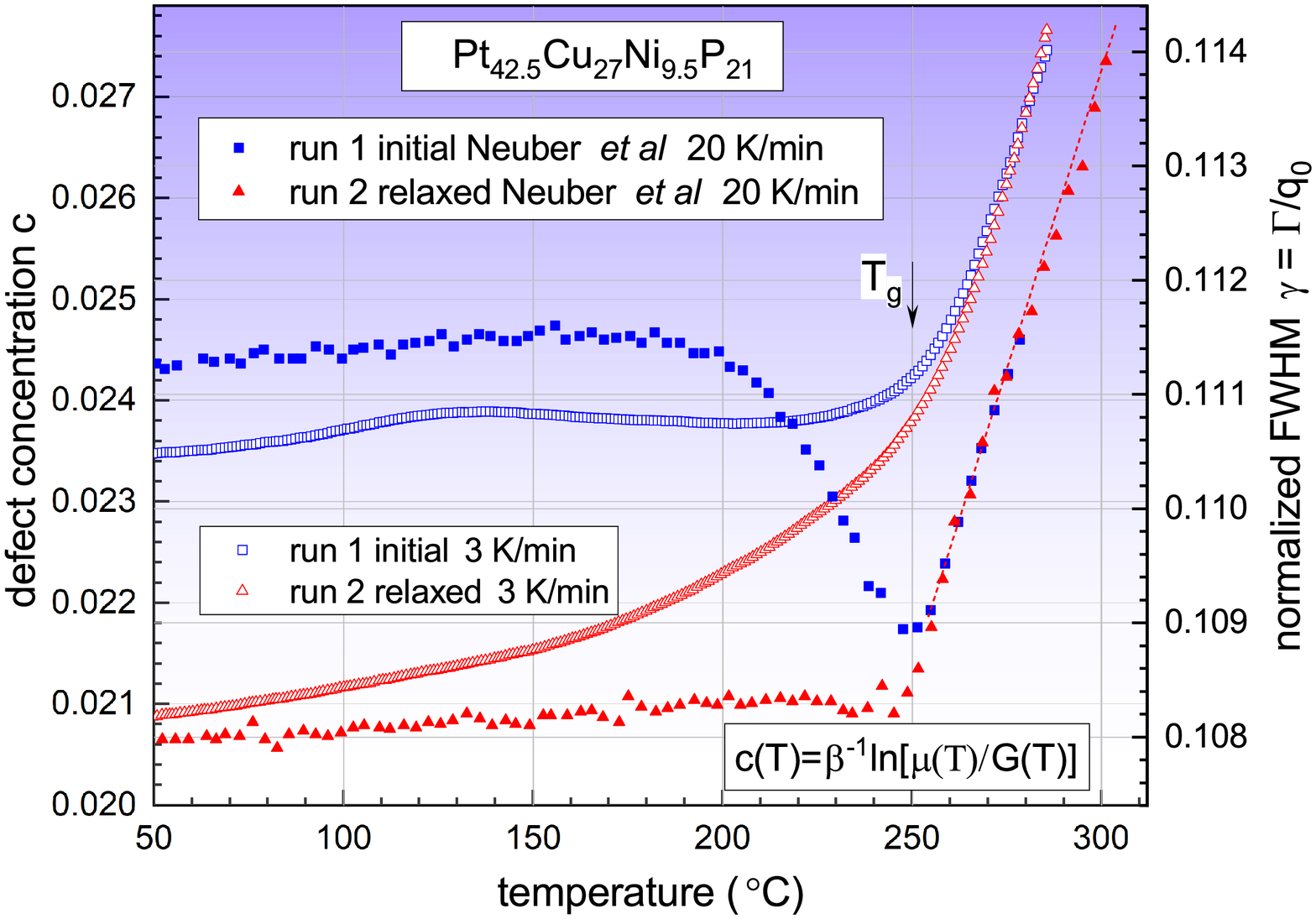}
\caption[*]{\label{Fig4.eps} Temperature dependences of the defect concentration in the initial state (run 1) and after the relaxation (run 2) calculated using Eq.(\ref{c}) for a heating rate of 3 K/min. The Figure also shows temperature dependences of the normalized FWHM $\gamma=\Gamma/q_0$, where the FWHM $\Gamma$ is taken from Neuber \textit{et al.} work \cite{NeuberActaMater2021} for a heating rate of 20 K/min and $q_0$ is the scattering vector corresponding to the first $S(q)$ peak. The calorimetric glass transition temperature $T_g$ determined at 3 K/min is indicated. The scales in left and right ordinate axes are equal and, therefore, $c$ is proportional to $\gamma$ in the supercooled liquid state (i.e. at $T>T_g$).  }
\end{center}
\end{figure} 

One can now calculate the defect concentration (\ref{c}) assuming $\alpha\approx 1$ as mentioned above. Temperature dependence $c(T)$ for the initial (run 1) and relaxed (run 2) states is given in Fig.\ref{Fig4.eps} for the heating rate of 3 K/min. In the initial state, $c$ weakly depends on temperature up to $T_g$ but rapidly increases with $T$ in the supercooled liquid state above $T_g$. After the relaxation (run 2), the defect concentration at room temperature becomes smaller by $\approx 13\%$ and moderately increases with temperature upon subsequent heating below $T_g$. Near $T_g$, the concentration starts to rapidly increase so that in the supercooled liquid state temperature dependence $c(T)$ in the relaxed state  coincides with that for the initial state. Heating into the supercooled liquid state, therefore, completely removes the "memory" of the preceding thermal prehistory.

For further analysis, we used the FWHM data for the same glass derived from high-energy synchrotron X-ray diffraction as a function of temperature reported by Neuber \textit{et al} \cite{NeuberActaMater2021}. The original FWHM data from this work (see SI Fig.6 in Ref.\cite{NeuberActaMater2021}) were transformed into the \textit{normalized} FWHM as $\gamma=\Gamma/q_0$, where $\Gamma$ is the absolute FWHM of the first $S(q)$ peak given in Ref.\cite{NeuberActaMater2021} and $q_0 = 2.9$  $\textup{~\AA}^{-1}$ is the scattering vector corresponding to this peak \cite{GrossCommnPhys2019} (the same $q_0$-value comes from our X-ray data shown in Fig.\ref{Fig1.eps}).

Temperature dependence of the normalized FWHM $\gamma$ is given in Fig.\ref{Fig4.eps} (X-ray diffraction patterns were measured at a rate of 20 K/min \cite{NeuberActaMater2021}) together with temperature dependence of the defect  concentration $c$. It is seen that $\gamma$ in the initial state (run 1) is almost temperature independent up to a temperature $T\approx 200 ^\circ$C. At higher temperatures  $200 ^\circ$C$<T<T_g\approx 250^\circ$C, $\gamma$ decreases by $\approx 2\%$ that is ascribed to structural relaxation \cite{NeuberActaMater2021}. Upon further heating above $T_g$, the normalized FWHM rapidly increases with temperature. After the relaxation produced by heating up to $278^\circ$C and cooling to room temperature, $\gamma$ decreases by $\approx 3\%$. Upon subsequent heating (run 2), $\gamma$ is nearly temperature independent up to $T_g$. However, in the supercooled liquid region above $T_g$, the normalized FWHM rapidly increases with temperature coinciding  with that during run 1. It should be emphasized that quite similar behavior of the FWHM   upon heating/cooling of a high-entropy metallic glass below $T_g$ was recently reported by Luan \textit{et al}. \cite{LuanNatCommun2022}.

\begin{figure}[t]
\begin{center}
\includegraphics[scale=1.1]{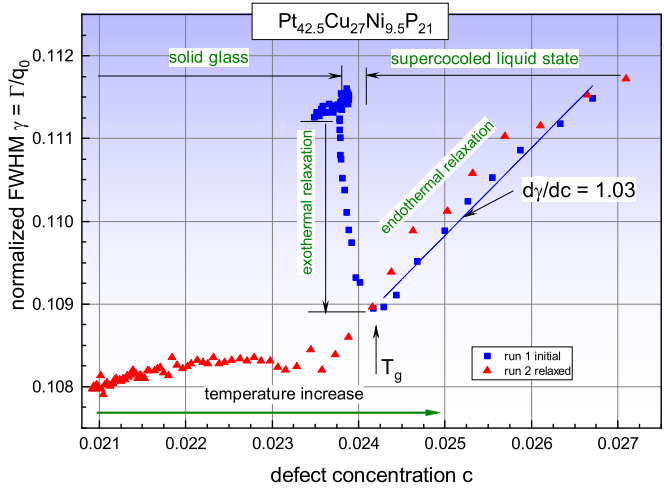}
\caption[*]{\label{Fig5.eps} Dependence of the normalized FWHM $\gamma$ on the defect concentration $c$ during run 1 (initial state) and subsequent run 2 (relaxed state). Glass transition temperature $T_g$ at 3 K/min and the direction of temperature increase are shown by the arrows. It is seen that $\gamma (c)$-dependences  below $T_g$  are completely different for the initial and relaxed states while $\gamma(c)$ for these states in the superccoled liquid region ($T>T_g$) nearly coincide. }
\end{center}
\end{figure} 

A comparison of temperature dependences of the defect concentration $c$ and normalized FWHM $\gamma$ given in Fig.\ref{Fig4.eps} clearly shows that they are similar. Below $T_g$, the concentration $c$ and $\gamma$ weakly depend on temperature although  some minor differences are observed. Above $T_g$, both quantities linearly increase with temperature while the corresponding derivatives are very close, $dc/dT \approx (1.12\pm 0.06)\times 10^{-4}$ and $\gamma/dT\approx (1.00 \pm 0.02)\times 10^{-4}$. The relaxation-induced changes of these quantities are also close, $\Delta c\approx 0.0026$ and $\Delta \gamma\approx 0.0030$ (calculated as corresponding differences for runs 1 and 2 at $50^\circ$C). The most essential difference between these quantities is that the absolute $\gamma$-values are by about 0.09 bigger than the absolute values of $c$. Possible reasons for this difference are discussed below.

A better view of the relationship between $\gamma$ and $c$ can be obtained if temperature in Fig.\ref{Fig4.eps} is excluded that provides a direct view of $\gamma (c)$-function. This is done in Fig.\ref{Fig5.eps}, which shows that the behavior of  $\gamma (c)$ is completely different below and above $T_g$. In the initial state (run 1), $c$ and $\gamma$ are nearly constant that corresponds to the temperature range $30^\circ$C$ <T< 240^\circ$C in Fig.\ref{Fig4.eps}. Further heating of the initial sample results in a quick $\gamma$-drop in a very narrow range of the concentrations. At $T>T_g$, $\gamma$ linearly increases with $c$. The relaxed sample (run 2) displays nearly no $\gamma$ changes upon a significant increase of $c$ occurring upon heating to $\approx T_g$. Above $T_g$, $\gamma$ linearly increases with $c$ just in the same way as for the initial state.

The following conclusions can be drawn from the data given in Figs \ref{Fig4.eps} and \ref{Fig5.eps}. Heating of the initial samples up to $200^\circ$C  almost does not  change $\gamma$ while $c$ is also nearly constant, which is quite natural for a "frozen" glass structure without any relaxation. However, fast "relaxation" (see Fig.\ref{Fig5.eps}) takes place at almost constant defect concentration while DSC shows significant exothermal reaction (Fig.\ref{Fig2.eps}).  Above $T_g$, the defect concentration rapidly increases with temperature leading to an increase of the amount of disordered regions of structure, which is directly reflected in a rise of the normalized FWHM.

The defect concentration below $T_g$ is strongly reduced in the relaxed state (run 2) as  seen in Figs \ref{Fig4.eps} and \ref{Fig5.eps}. Shear modulus measurements in this case indicate a monotonous increase of the defect concentration (Fig.\ref{Fig4.eps}) and the presence of a notable endothermal effect (Fig.\ref{Fig2.eps}) while $\gamma$ remains nearly unchanged. The $\gamma (c)$-dependence for the relaxed state above $T_g$ is exactly the same as that for the initial samples: $\gamma$ linearly increases with $c$ with the same slope           
$d\gamma/dc\approx 1.03$. Taking into account strong heat absorption  (Fig.\ref{Fig2.eps}) and  shear softening above $T_g$ (Fig.\ref{Fig3.eps}) one can conclude that the supercooled liquid state is characterized by the intense defect multiplication (Fig.\ref{Fig4.eps}), which significantly destroys  the dominant short-range order and leads to a large $\gamma$-increase. Since the memory of the thermal prehistory is lost above $T_g$, this behavior takes place for both initial and relaxed samples.   Thus, changes in the defect subsystem of glass can lead to either defect-induced disordering (increasing $\gamma$) or ordering (decreasing $\Gamma$) depending on temperature and thermal prehistory. Possible reasons for the behaviors of $c$ and $\gamma$ are considered below.

It is should be also once more pointed out that the defect concentration was calculated for the heating rate of 3 K/min while the FWHM was measured at 20 K/min. A more correct analysis of the relation between these quantities should be carried out using the data taken at the same heating rate. Due to technical limitations, shear modulus measurements with the present EMAT technique at 20 K/min are not possible. However, we tried to estimate what could happen if $G(T)$ measurements would have been performed at 20 K/min. For this, the $G(T)$-function taken at 3 K/min was shifted by 12 K towards higher temperatures according to the $T_g$-difference at the above rates (see Fig.\ref{Fig2.eps}c). On the other hand, we also tried to shift the $\gamma(T)$-data towards lower temperatures by the same quantity modeling thus a decrease of the heating rate upon X-ray measurements. Both procedures result in some minor changes of the $\gamma(c)$-plot but the derivative $d\gamma/dc$ in the supecooled liquid state remains almost the same.            
 
\subsection{Molecular dynamic simulation}

Model Fe$_{20}$Ni$_{20}$Cr$_{20}$Co$_{20}$Cu$_{20}$ melt was quenched to $T=0$ K at different rates as indicated below. An increase of the quenching rate results in pronounced shear softening that can be rationalized in terms of an increase of the defect concentration. The identification of defect regions and determination  of their concentration can be  done  qualitatively by the method presented in Refs \cite{KretovaJETPLett2020,KonchakovJETPLett2022}. For quantitative estimates of 
the defect concentration, we applied Eq.(\ref{c}) derived within the framework of the IT, as discussed above. The shear moduli of glass and maternal FCC crystal were determined by applying a small shear strain  of about $10^{-3}$. Then, the defect concentration was calculated using Eq.(\ref{c}) with the product $\alpha \beta=15.4$ as determined earlier \cite{KretovaJETPLett2020}.     

\begin{figure}[t]
\begin{center}
\includegraphics[scale=0.35]{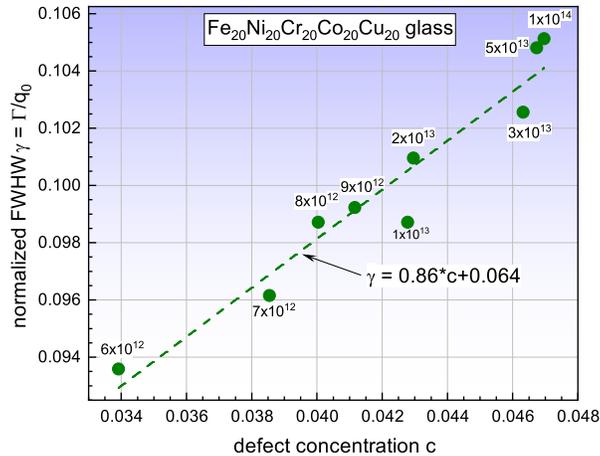}
\caption[*]{\label{Fig6.eps} Dependence of the normalized FWHM $\gamma$ on the defect concentration $c$  calculated using Eq.(\ref{c}) for model glassy Fe$_{20}$Ni$_{20}$Cr$_{20}$Co$_{20}$Cu$_{20}$ produced with different melt quenching rates (indicated in K/s). The straight line gives a least-square linear fit.  }
\end{center}
\end{figure} 

Figure \ref{Fig6.eps} shows the normalized FWHM $\gamma$ as a function of the defect concentration $c$ where the data points correspond to the melt quenching rates  indicated in K/s. It is seen that $\gamma$ linearly increases with $c$ just as in the case of experimental data for Pt-based glass for temperatures $T>T_g$ while the absolute $\gamma$-values are close to those given in Ref.\cite{NeuberActaMater2021}.  At that, the  slope of $\gamma (c)$-dependence, $d\gamma/dc=0.86$, which is fairly close to the experiment on Pt-based glass (Fig.\ref{Fig5.eps}).  In general, one can conclude that the simulation supports the experiment giving similar results.

\subsection{Comments on the relation between the defect concentration and FWHM }

Let us consider possible reasons for the differences in $c$ and $\gamma$  absolute values. It should be first of all noted noted that the defect concentration $c$ in Eq.(\ref{c}) depends on the choice of the shear susceptibility $\beta$, which is determined to a precision of $10-15\%$. Besides that, the above calculation with this equation for the Pt-based glass  assumed $\alpha=1$ since the exact value of this quantity is unknown. The estimates for real MGs give $\alpha \approx 0.6$  while molecular static simulations of crystalline pure metals and  Fe$_{20}$Ni$_{20}$Cr$_{20}$Co$_{20}$Cu$_{20}$  lead to the values $0.50\leq \alpha \leq 0.78$ \cite{MakarovJNCS2021}. The uncertainties in  the choice of $\alpha$ and $\beta$  will lead to  corresponding changes of the defect concentration and, therefore, to a shift of the abscissas in Figs \ref{Fig5.eps} and \ref{Fig6.eps}. However, such redefining of $c$  does not generally affect the consideration given above. 

It is to be also noted that the preparation and heat treatment of samples in the present work and Ref.\cite{NeuberActaMater2021} are notably different. In the latter case, quenched samples for the X-ray investigation were relatively thin (0.5 mm thick) while the castings in the present work are 2 mm thick. Therefore, the melt quenching rate in our work is notably smaller and the samples are more relaxed. Moreover, the relaxed state (i.e. run 2) of the samples in Ref.\cite{NeuberActaMater2021} were obtained by heating and cooling at 20 K/min while the  rate of 3 K/min was applied in the present work. Thus, the defect concentration  was the largest in the initial samples in Ref.\cite{NeuberActaMater2021} and minimal in the relaxed specimens tested in our work. The initial samples in the present work and relaxed samples in Ref.\cite{NeuberActaMater2021} should have some intermediate defect concentration. This is in a qualitative agreement with temperature dependences of $c$ and $\gamma$ below $T_g$. Indeed, temperature dependence of $\gamma$ for the initial samples in Ref.\cite{NeuberActaMater2021} shows an abrupt decrease near $T_g$ (Fig.\ref{Fig4.eps}). At that, $c$ and $\gamma$ for the relaxed specimens tested in Ref.\cite{NeuberActaMater2021} and initial samples in the present work are almost temperature independent up to $T_g$ but rapidly increase at higher temperatures. The relaxed samples in our work demonstrate a notable growth of the defect concentration even below $T_g$.

Another possible reason in the difference of the absolute values of $c$ and $\gamma$ could be related with additional sources of $S(q)$ broadening. In part, this possibility is supported by the fact that the extrapolation of  $\gamma (c)$-dependence in Fig.\ref{Fig6.eps} towards small $c$ does not lead to zero $\gamma$-values but approaches a constant of about $\approx 0.064$, which is close to the difference between the absolute values of $\gamma$ and $c$ in Fig.\ref{Fig4.eps}. It is also to be noted that the extrapolation $\gamma (c)$-dependence in Fig.\ref{Fig5.eps} similarly does not lead  to zero $\gamma$-values  upon $c\rightarrow 0$ although this is less pronounced as compared with Fig.\ref{Fig6.eps}.   The reason for additional $S(q)$ broadening should be evidently related to the non-crystalline  "matrix" itself, which, as mentioned above, can be considered as an array of relatively large clusters with predominant (e.g. icosahedral) short range order. However, the investigation described above indicates that the defects provide larger $S(q)$ broadening at $T>T_g$. This investigation provides a simple measure of defect-induced ordering/disordering in the supercooled liquid state since the changes between the normalized FWHM and defect concentration are nearly equal, i.e.  $\Delta\gamma \approx \Delta c$.

\section{Conclusions}

We performed precise measurements of the high-frequency shear modulus of bulk Pt$_{42.5}$Cu$_{27}$Ni$_{9.5}$P$_{21}$ glass in the initial and relaxed (preannealed) states  and on this basis calculated temperature dependence of the defect concentration  using the Interstitialcy theory. These calculations were compared with temperature dependence of the full width at half maximum (FWHM) $\Gamma$ of the first $S(q)$-peak derived by Neuber \textit{et al}. \cite{NeuberActaMater2021} using synchrotron X-ray diffraction data taken on  the same glass. 

It is found that the normalized FWHM $\gamma=\Gamma/q_0$ ($q_0$ is the scattering vector corresponding to the first $S(q)$ peak), which constitutes an integral measure of structural disordering, linearly increases with the defect concentration $c$ above the glass transition temperature $T_g$ in the same way both for initial and relaxed  specimens.  This means that strong heat absorption  and related rapid $c$-increase  in the supercooled liquid region provide a significant defect-induced disruption of the dominant short-range order increasing thus the integral structural disordering. The determined fact that $\Delta \gamma\approx \Delta c$ in this region provides a simple quantitative measure of this disordering. 

Below $T_g$, the interrelation between $\gamma$ and $c$ is quite complicated and entirely  different for  initial and relaxed samples. In the former case, strong defect-induced ordering upon approaching $T_g$ is observed while relaxed samples do not reveal any clear ordering/disordering. Possible reasons for these differences can be due to the variations in the defect concentrations because of different degree of relaxation of specimens used for the measurements of $c$ and $\gamma$ as well as the presence of additional sources of $S(q)$ broadening, which are not related to the defect subsystem.

We also performed molecular dynamic simulation of the relationship between $\gamma$ and $c$. Since the interatomic interaction in the  Pt-based glass under investigation is unknown, we performed modeling of high-entropy non-crystalline Fe$_{20}$Ni$_{20}$Cr$_{20}$Co$_{20}$Cu$_{20}$. The defect concentration was again calculated using the Interstitialcy theory. It was found that $\gamma$ linearly increases with $c$ in the same way as in the case of Pt-based glass. At that, the derivative $d\gamma /dc$  is also close to unity. Overall, the simulation agrees with the results obtained by the analysis of experimental data on Pt-based glass. 

\section*{Acknowledgements}

The work was supported by the Russian Science Foundation under the project 20-62-46003. The kind help of Prof. A.S. Aronin (Institute for Solid State Physics RAS) with X-ray measurements is greatly acknowledged.


\begin{thebibliography}{999}

\bibitem{ChengProgMatSci2011} Y.Q. Cheng, E. Ma, Atomic-level structure and structure-property relationship in metallic glasses, Prog. Mater. Sci. 56 (2011) 379-473.

\bibitem{NeuberActaMater2021} N. Neuber, O. Gross, M. Frey, B. Bochtler, A. Kuball, S. Hechler, I. Gallino, R. Busch, On the thermodynamics and its connection to structure in the Pt-Pd-Cu-Ni-P bulk metallic glass forming system, Acta Mater 220 (2021) 117300. 

\bibitem{YavariActaMater2005} A.R. Yavari, A.L. Moulec, A. Inoue, N. Nishiyama, N. Lupu, E. Matsubara, W.J. Botta, G. Vaughan, M.D. Michiel, A. Kvick, Excess free volume in metallic glasses measured by X-ray diffraction, Acta Mater. 53 (2005) 1611-1619.

\bibitem{GeorgarakisActaMater2015}  K. Georgarakis, L. Hennet, G.A. Evangelakis, J. Antonowicz, G.B. Bokas, V. Honkimaki, A. Bytchkov, M.W. Chen, A.R. Yavari, Acta Mater. 87 (2015) 174–186.

\bibitem{SinghJALCOM2015}  D. Singh, R.K. Mandal, R.S. Tiwari, O.N. Srivastava, Effect of cooling rate on the crystallization and mechanical behaviour of Zr-Ga-Cu-Ni metallic glass composition, J. Alloys Comp. 648 (2015) 456-462. 

\bibitem{DuMaterToday2020} Q. Du, X. Liu, H. Fan, Q. Zeng, Y. Wu, H. Wang, D. Chatterjee, Y. Ren, Y. Ke, P.M. Voyles, Z. Lu, E. Ma, Reentrant glass transition leading to ultrastable metallic glass, Mater. Today 34 (2020) 66-77.

\bibitem{LuanNatCommun2022} H. Luan,10, X. Zhang, H. Ding, F. Zhang, J.H. Luan, Z.B. Jiao, Y.-C. Yang, H. Bu, R. Wang, J. Gu, C. Shao, Q. Yu, Y. Shao, Q. Zeng, N. Chen, C.T. Liu, K.-F. Yao, High-entropy induced a glass-to-glass transition in a metallic glass, Nature Commun. 13 (2022) 2183.

\bibitem{ChenMaterSciEng2008} N. Chen, K.-F. Yao, F. Ruan, Y.-Q. Yang, The influence of cooling rate on the hardness of Pd–Si binary glassy alloys, Mater. Sci. Eng. A 473 (2008) 274–278.

\bibitem{HuangMaterDes2014} Y. Huang, H. Fan, D. Wang, Y. Sun, F. Liu, J. Shen, J. Sun, J. Mi, The effect of cooling rate on the wear performance of a ZrCuAlAg bulk metallic glass, Mater. Design 58 (2014) 284–289.

\bibitem{XuMaterSciEng2015} Y. Xu,B. Shi, Z. Ma,J. Li Evolution of shear bands, free volume, and structure in room temperature rolled Pd$_{40}$Ni$_{40}$P$_{20}$ bulk metallic glass, Mater. Sci. Eng. A 623 (2015) 145-152.

\bibitem{AdachiMaterrSciEng2015} N. Adachi, Y. Todaka, Y. koyama, M. Umemoto, Cause of hardening and softening in the bulk glassy alloy Zr$_{50}$Cu$_{40}$Al$_{10}$ after high-pressure torsion, Mater. Sci. Eng. A 627(2015) 171–181.

\bibitem{CaoActaMater2006} Q.P. Cao, J.F. Li, Y.H. Zhou, A. Horsewell, J.Z. Jiang,  Effect of rolling deformation on the microstructure of bulk Cu$_{60}$Zr$_{20}$Ti$_{20}$ metallic glass and its crystallization, Acta Mater. 54 (2006) 4373–4383.

\bibitem{EbnerActaMater2018} C. Ebner, B. Escher, C. Gammer, J. Eckert, S. Pauly, C. Rentenberger, Structural and mechanical characterization of heterogeneities in a CuZr-based bulk metallic glass processed by high pressure torsion, Acta Mater. 160 (2018) 147-157.

\bibitem{GunderovMaterLett2020} D.V. Gunderov, A.A. Churakova, V.V. Astanin, R.N. Asfandiyarov, H. Hahn, R.Z. Valiev, Accumulative HPT of Zr-based bulk metallic glasses, Mater. Lett. 261 (2020) 127000.

\bibitem{LiNatureMater2022} M.-X. Li, Y.-T. Sun, C. Wang, L.-W. Hu, S. Sohn, J. Schroers,
W.-H. Wang,  Y.-H. Liu, Data-driven discovery of a universal indicator for
metallic glass forming ability, Nature Mater 21 (2022) 165-172.

\bibitem{HirataScience2013} A. Hirata, L.J. Kang, T. Fujita, B. Klumov, K. Matsue, M. Kotani, A.R. Yavari, M.W. Chen, Geometric frustration of icosahedron in metallic glasses, Science 341  (2013) 376-379.

\bibitem{BalakirevRevSciInstrum2019} F.F. Balakirev, S.M. Ennaceur, R. J. Migliori, B. Maiorov, A. Migliori, Resonant ultrasound spectroscopy: the essential toolbox, Rev. Sci. Instrum. 90 (2019) 121401.

\bibitem{VasilievUFN1983} A.N. Vasil'ev, Yu.P. Gaidukov, Electromagnetic excitation of sound in metals, Sov. Phys. Usp. 26 (1983) 952-973. 

\bibitem{PlimptonJCompPys1995} S. Plimpton, Fast parallel algorithms for short-range molecular dynamics, J. Comp. Phys. 117 (1995) 1-19.

\bibitem{FarkasJMaterRes2018} D. Farkas, A.Caro,  Model interatomic potentials and lattice strain in a high-entropy alloy, J. Mater. Res. 33 (2018) 3218–3225.

\bibitem{KretovaJETPLett2020} M.A. Kretova, R.A. Konchakov, N.P. Kobelev, V.A.Khonik,  Point defects and their properties in the Fe20Ni20Cr20Co20Cu20 high-entropy Alloy, J. Exp. Theor.Phys. Lett. 111 (2020) 679–684.    

\bibitem{KonchakovJETPLett2022} R.A. Konchakov, A.S. Makarov, A.S. Aronin, N.P. Kobelev, V.A. Khonik, Elastic dipoles in crystal and glassy aluminum and high-entropy Fe20Ni20Cr20Co20Cu20 alloy, J. Exp. Theor.Phys. Lett.  115 (2022) 280–285.

\bibitem{StukowskiModelSimulMaterSciEng2010} A. Stukowski,  Visualization and analysis of atomistic simulation data with OVITO – the Open Visualization Tool,  Model. Simul. Mater. Sci. Eng. 18 (2010) 015012.

\bibitem{KhonikMetals2019}	V.A. Khonik, N. Kobelev. Metallic glasses: a new approach to the understanding of the defect structure and physical properties. Metals 9 (2019) 605. 

\bibitem{MakarovJPCM2021} A.S. Makarov, J.C. Qiao, N.P. Kobelev,  A.S. Aronin,  V.A. Khonik, Relation of the fragility and heat capacity jump in the supercooled liquid region with the shear modulus relaxation in metallic glasses, J. Phys. Cond. Matter 33 (2021) 275701.

\bibitem{MakarovJETPLett2022} A.S. Makarov, M.A. Kretova, G. V. Afonin, J.C. Qiao, A.M. Glezer, N.P. Kobelev, V.A. Khonik. On the nature of the excess internal energy and entropy of metallic glasses. J. Exp. Theor. Phys. Lett. 115 (2022) 102-107.

\bibitem{MakarovIntermetallics2017}  A.S. Makarov, Yu.P. Mitrofanov, G.V. Afonin, N.P. Kobelev, V.A. Khonik, Shear susceptibility - a universal integral parameter relating the shear softening, heat effects, anharmonicity of interatomic interaction and "defect" structure of metallic glasses. Intermetallics 87 (2017) 1-5.
 
\bibitem{MitrofanovJALCOM2016}  Yu.P. Mitrofanov, D.P. Wang, W.H. Wang, V.A. Khonik, Interrelationship between heat release and shear modulus change due to structural relaxation of bulk metallic glasses, J. Alloys Comp. 677 (2016) 80-86.

\bibitem{GrossCommnPhys2019}  O. Gross, N. Neuber, A. Kuball, B. Bochtler, S. Hechler, M. Frey, R. Busch, Commun. Phys. 2 (2019) 83.

\bibitem{AfoninJETP2020} G.V. Afonin, Yu.P. Mitrofanov, N.P. Kobelev, V.A. Khonik,  Shear modulus relaxation and thermal effects in a Zr$_{65}$Cu$_{15}$Ni$_{10}$Al$_{10}$ metallic glass after inhomogeneous plastic deformation. J. Exp. Theor. Phys.  131 (2020) 582-588.

\bibitem{MakarovJNCS2021} A.S. Makarov, G.V. Afonin, R.A. Konchakov, J.C. Qiao, A.S. Aronin, N.P. Kobelev, V.A. Khonik, One-to-one correlation between the kinetics of the enthalpy changes and the number of defects assumed responsible for structural relaxation in metallic glasses, J. Non-Cryst. Sol. 558 (2021) 120672.


\end{thebibliography}
\end{document}